\documentclass[12pt, draftclsnofoot, onecolumn]{IEEEtran}%Single Column
\usepackage{enumerate}
\usepackage{cite}
\usepackage{float}
\usepackage{mathrsfs}
\usepackage{color}
\usepackage{caption}
\makeatletter

\newcommand{\Rmnum}[1]{\expandafter\@slowromancap\romannumeral #1@}

\makeatother
\usepackage{multirow}
\usepackage{algorithm}
\usepackage{algpseudocode}

\usepackage{graphicx}%zhou
\ifCLASSINFOpdf

\else

\fi

\usepackage[cmex10]{amsmath}
\usepackage{bm}%zhou

\usepackage{epstopdf}

\usepackage{cite}
\usepackage[cmex10]{amsmath}
\usepackage{array}
\usepackage{psfrag}
\usepackage{subfigure}
\usepackage{epsfig}
\usepackage{amssymb}

\usepackage{url}

\usepackage{booktabs} %used for table

\begin{document}
\title{Sparse Code Multiple Access for 6G Wireless Communication Networks: Recent Advances and Future Directions}

%%Author
\author{Lisu Yu, \IEEEmembership{Member, IEEE,}
        Zilong Liu, \IEEEmembership{Senior Member, IEEE,}
        Miaowen Wen, \IEEEmembership{Senior Member, IEEE,}
        Donghong Cai, \IEEEmembership{Member, IEEE,} 
        Shuping Dang, \IEEEmembership{Member, IEEE,}
        Yuhao Wang, \IEEEmembership{Senior Member, IEEE,}
        and Pei Xiao, \IEEEmembership{Senior Member, IEEE}
\thanks{Copyright (c) 2021 IEEE. Personal use of this material is permitted. However, permission to use this material for any other purposes must be obtained from the IEEE by sending a request to pubs-permissions@ieee.org.}
%\thanks{The work of L. Yu was supported in part by the State Key Laboratory of Computer Architecture (ICT, CAS) Open Project under Grant CARCHB202019, the National Natural Science Foundation of China under Grants 41865002, 61761030, and 62001201, and the Natural Science Foundation of Jiangxi under Grant 20171BAB202007. 
%The work of Z. Liu and P. Xiao was supported in part by the UK Engineering and Physical Sciences Research Council under Grant EP/P03456X/1.
%The work of M. Wen was supported in part by the Fundamental Research Funds for the Central Universities under Grant 2019SJ02. The work of D. Cai was supported in part by the National Natural Science Foundation of China (NSFC) under Grant 62001190. The work of Y. Wang was supported in part by the National Natural Science Foundation of China under Grant 62061030, the National Key Research and Development Project under Grants 2018YFB1404303 and 2018YFB14043033. The corresponding authors are Miaowen Wen and Yuhao Wang.}
\thanks{L. Yu is with the School of Information Engineering, Nanchang University, Nanchang 330031, China, also with the State Key Laboratory of Computer
Architecture, Institute of Computing Technology, Chinese Academy of Sciences, Beijing 100190, China (e-mail: lisuyu@ncu.edu.cn).}
\thanks{Z. Liu is with School of Computer Science and Electronics Engineering, University of Essex, Colchester CO4 3SQ, U.K. (e-mail: zilong.liu@essex.ac.uk).}
\thanks{M. Wen is with the School of Electronic and Information Engineering, South China University of Technology, Guangzhou 510640, China (e-mail: eemwwen@scut.edu.cn).}
\thanks{D. Cai is with the College of Cyber Security, Jinan University, Guangzhou 510632, China (e-mail: dhcai@jnu.edu.cn).}
\thanks{S. Dang is with Computer, Electrical and Mathematical Science and Engineering Division, King Abdullah University of Science and Technology (KAUST), Thuwal 23955-6900, Saudi Arabia (e-mail: shuping.dang@kaust.edu.sa).}
\thanks{Y. Wang is with the School of Information Engineering, Nanchang University, Nanchang 330031, P. R. China (e-mail: wangyuhao@ncu.edu.cn).}
\thanks{P. Xiao is with Institute for Communication Systems, 5G Innovation Centre, University of Surrey, Guildford GU2 7XH, U.K. (e-mail: p.xiao@surrey.ac.uk).}
}

\maketitle

\begin{abstract}
%\boldmath
As 5G networks rolling out in many different countries nowadays, the time has come to investigate how to upgrade and expand them towards 6G, where the latter is expected to realize the interconnection of everything as well as the development of a ubiquitous intelligent mobile world for intelligent life. 
To enable this epic leap in communications, this article provides an overview and outlook on the
application of sparse code multiple access (SCMA) for 6G wireless communication systems, which is an emerging disruptive non-orthogonal multiple access (NOMA) scheme for the enabling of massive connectivity. We propose to apply SCMA to a massively distributed access system (MDAS), whose architecture is based on fiber-based visible light communication (FVLC), ultra-dense network (UDN), and NOMA. Under this framework, we consider the interactions between optical front-hauls and wireless access links. In order to stimulate more upcoming research in this area, we outline a number of promising directions associated with SCMA for faster, more reliable, and more efficient multiple access in future 6G communication networks.
\end{abstract}


\IEEEpeerreviewmaketitle

\section{Introduction}

\IEEEPARstart{W}{ireless} communication has been considered as one of the most successful technological innovations in the 21st century. Every new generation mobile communication emerges roughly every 10 years since 1980s. In retrospect, the first generation (1G) mobile communication started with analog voice using advanced mobile phone system (AMPS). Around 1990, digital voices were transmitted using global system for mobile communications (GSM), which forms a key component of second generation (2G) systems. Mobile broadband became available around 2000, thanks to the use of code division multiple access (CDMA) technology, which is a core technology of third generation (3G) mobile communication. Significant boost of data rates has been achieved in fourth generation (4G) mobile communication era driven by the strong demand for multimedia communications. Nowadays, fifth-generation (5G) has become commercially available and is ready for wide deployment.  

The last five years have witnessed a tremendous development in cellular networks towards data-oriented services that include, but are not limited to, multimedia, online gaming, and high-quality video streaming. As a result, the number of mobile subscribers and the amount of data traffic have increased explosively \cite{agiwal2016next}. The advent of internet of things (IoT) and machine-type communication networks has imposed further requirements for higher wireless network capacity. These challenges are considered to be a critical driver pushing the 5G wireless communication networks.

With the explosive growth of ubiquitous mobile services, the mobile communication systems are facing formidable challenges imposed by the need for a large number of concurrent services with extremely high connection density. Time has come to investigate the next generation wireless communication systems, such as beyond 5G (B5G) and the sixth generation (6G) mobile communication systems. One of the most important technologies in every generation of mobile communication systems is the multiple access (MA) scheme, which is one of the most important techniques in wireless communications. So far, we have adopted frequency-division multiple access (FDMA, 1G), time-division multiple access (TDMA, 2G), CDMA (3G), and orthogonal frequency-division multiple access (OFDMA, 4G). Although non-orthogonal multiple access (NOMA) has attracted extensive research attentions in 5G \cite{7973146}, it has not been adopted in 5G New Radio (5G NR) as no consensus has been achieved partly due to the tremendous responses from many major telecommunication companies. 

\begin{figure*}[!t]
\centering
\includegraphics[width=0.9\textwidth]{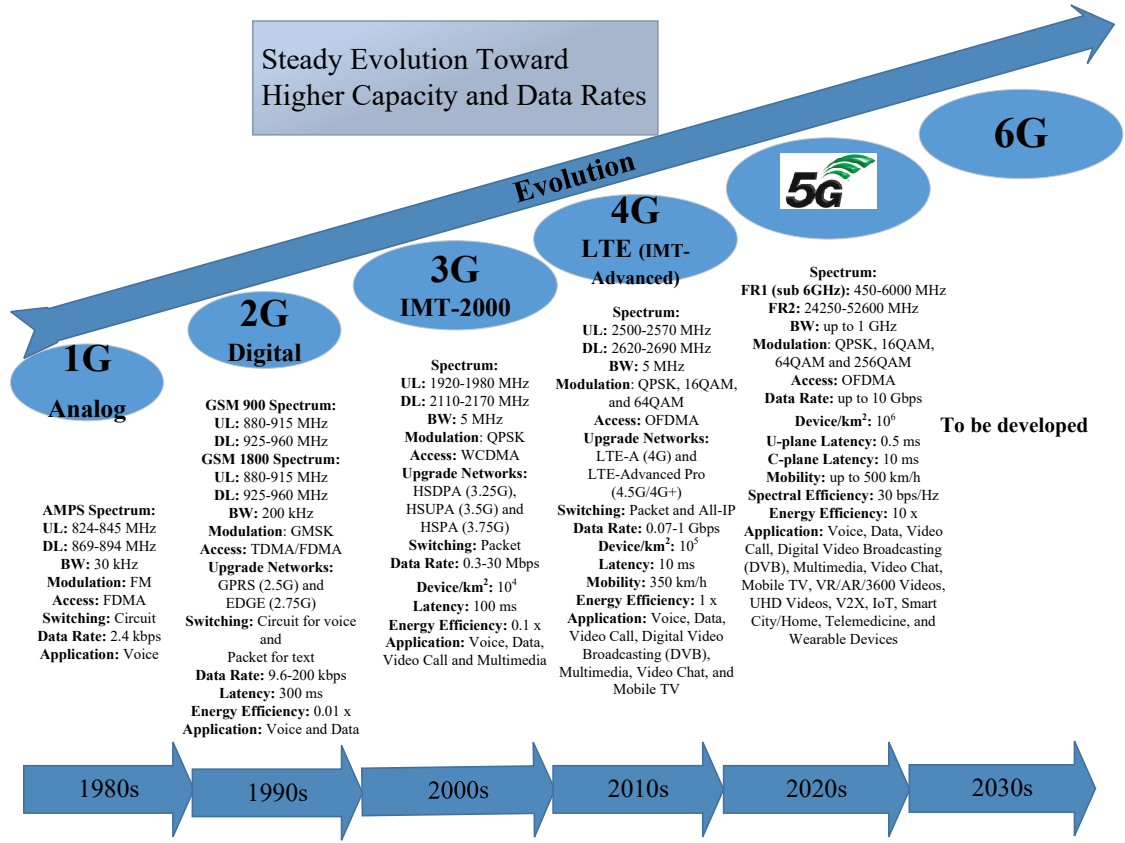}
\caption{Evolution of mobile networks.}\label{fig:6G}
\end{figure*}

Among them, sparse code multiple access (SCMA) is widely recognized a competitive candidate of code-domain NOMA scheme \cite{yu2018design}.
In SCMA, multiple users can be served simultaneously by employing different sparse codebooks. At the receiver, an efficient multi-user detector based on the message passing algorithm (MPA) is adopted, which explores the codebook sparsity to eliminate inter-user interference.
With 6G research starting to rise \cite{dang2020what}, it is time to ask how MA will evolve and then be adopted in 6G communication networks. In this work, we propose to use SCMA to support massively distributed access system (MDAS) in 6G for faster, more scalable, and more reliable, and more efficient massive access.

This article aims to provide a comprehensive overview and outlook on the SCMA assisted massive access for 6G wireless communication systems. We propose a new MDAS architecture based on fiber-based visible light communication (FVLC), ultra-dense network (UDN) \cite{kamel2016ultra} and NOMA. In this architecture, we consider the interactions between optical front-hauls and wireless access links. The optimum designs of MDAS need to be performed across the optical and wireless domains to jointly exploit the unique features and limits in both domains. We show that SCMA is an excellent candidate in shaping such a massive access system.

The remainder of this article is organized as follows. Section II provides an overview of the future perspective of 6G wireless communication networks. The proposed massively distributed system architecture is presented in Section III. Section IV introduces the details of sparse code multiple access, including the codebook design, performance analysis, and applications. Future directions are discussed in Section V. Section VI concludes the article.

\section{Versions of 6G Wireless Communication Networks}

This section presents our vision to the 6G wireless communication networks. To this end, we first provide a brief overview on mobile evolution from 1G to 5G. A summary of the major milestones of the current mobile networks can be found in Fig. \ref{fig:6G}. 

\subsection{Mobile Networks From 1G to 5G }
The first generation of mobile communication launched around 1980. Major subscribers were AMPS in North America, nordic mobile telephone (NMT) in Scandinavia, total access communication system (TACS) in the United Kingdom (UK) and Japan total access communications system (JTACS) in Japan. The first generation communication system was designed for voice communications with a data rate of up to 2.4 kbps, using a basic analog technology, in which frequency modulation (FM) and frequency division multiple access (FDMA) technology were adopted. However, The 1G system suffers from many  disadvantages, such as, low quality without encryption due to analog modulation, support of limited user equipments (UEs) due to FDMA technique, unreliable communications without a handoff process, and limited voice service.

In the early of 1990s, global systems for mobile communications (GSM) was adopted in 2G systems. A GSM system used digital techniques for voice communications with a data rate of up to 9.6 kbps, thanks to the usage of Gaussian minimum shift keying (GMSK) modulation and time division multiple access (TDMA) transmission technology with bandwidth (BW) of 200 kHz. The 2G systems benefited from the development of unified international standard for mobile communications, and can provide more services not limited to voice as well as enhanced the system capacity and security by digital encryption. To further improve the data rates of GSM, general packet radio services (GPRS) was developed which may be regarded as a 2.5G system with improved data rates up to 50 kbps using packet switching technology. GPRS represents an evolutionary step toward the enhanced data GSM environment (EDGE) system, which is considered as a pre-3G system. By employing eight-phase shift keying (8PSK) modulation technique along with GMSK, EDGE was designed to deliver data rates up to 200 kbps.

In 3G, high-speed Internet access has been introduced to greatly improve video and audio streaming capabilities using wideband code division multiple access (W-CDMA) and high speed packet access (HSPA) techniques. Specifically, HSPA has two mobile communication protocols, i.e., high speed downlink packet access (HSDPA) and high speed uplink packet access (HSUPA), which can further improve the performance of the traditional 3G mobile system. Aiming for evolved HSPA technology, an improved 3rd Generation Partnership Project (3GPP) standard, known as HSPA+ was released in late 2008. In 2010, the long term evolution (LTE) was introduced and subsequently named as 4G officially by International Telecommunication Union (ITU) and 3GPP.

In 4G systems, orthogonal frequency-division multiplexing (OFDM) technology has been employed to support scalable transmission bandwidths up to 20 MHz and advanced multi-antenna transmissions. Particularly, multiple-input multiple-output (MIMO) is a key technology that enables multi-stream transmissions for significantly higher spectrum efficiency. With LTE, peak mobile data rates up to 100 Mbps can be provided. To fulfill the higher capacity demand, LTE-Advanced (LTE-A) technology has been developed to support hundreds of megabits date rates. 
4G improves the networks' communication capability from the following respects: (i) network densification technique is deployed in areas with large numbers of UEs to improve network coverage and increase spectrum reuse; (ii) with spatial diversity, coordinated transmission/reception schemes and inter-cell interference cancellation solutions are adopted to reduce co-channel interference and enhance the spectral efficiency; (iii) LTE-A achieves bandwidth extensions (up to 100 MHz) via carrier aggregation to combine different component carriers, leading to enhanced spectrum utilization and higher network throughput.

Whilst the standardization of 5G systems is ongoing, 5G networks have been rolling out around the world. Based on the vision and requirements defined by ITU, 5G should satisfy three typical scenarios and eight key performance indexes (KPIs) \cite{chen2020vision}. Three scenarios include Gb/s-level data rates in enhanced mobile broadband (eMBB) systems, millisecond (ms) air interface latency in ultra-reliable low latency communication (URLLC) systems, and one million connections per square kilometer (1M/km$^2$) of massive machine type communication (mMTC) systems. To satisfy these KPIs, a series of enabling technologies were proposed, including massive MIMO, advanced coding and modulation, mmWave communication, NOMA, UDN, dual connectivity architecture, and so on. Moreover, emerging applications, such as virtual/augmented reality (VR/AR), autonomous driving, three-dimensional integrated communications are now pushing the step towards 6G. In these scenarios, we will need extremely higher data rates as well as hyper-fast communication access than what 5G networks will offer.

\subsection{6G Wireless Communication Networks}

5G networks may not be competent in fully supporting the development of a ubiquitous mobile society towards intelligent life. For example, in 6G networks, it is expected to achieve 10 to 100 times higher data rates and system capacity, higher spectrum efficiency and significantly lower access latency, compared with 5G networks \cite{chen2020vision}. It is envisaged that 6G will be characterized by the following features:

\begin{figure*}[!t]
\centering
\includegraphics[width=0.9\textwidth]{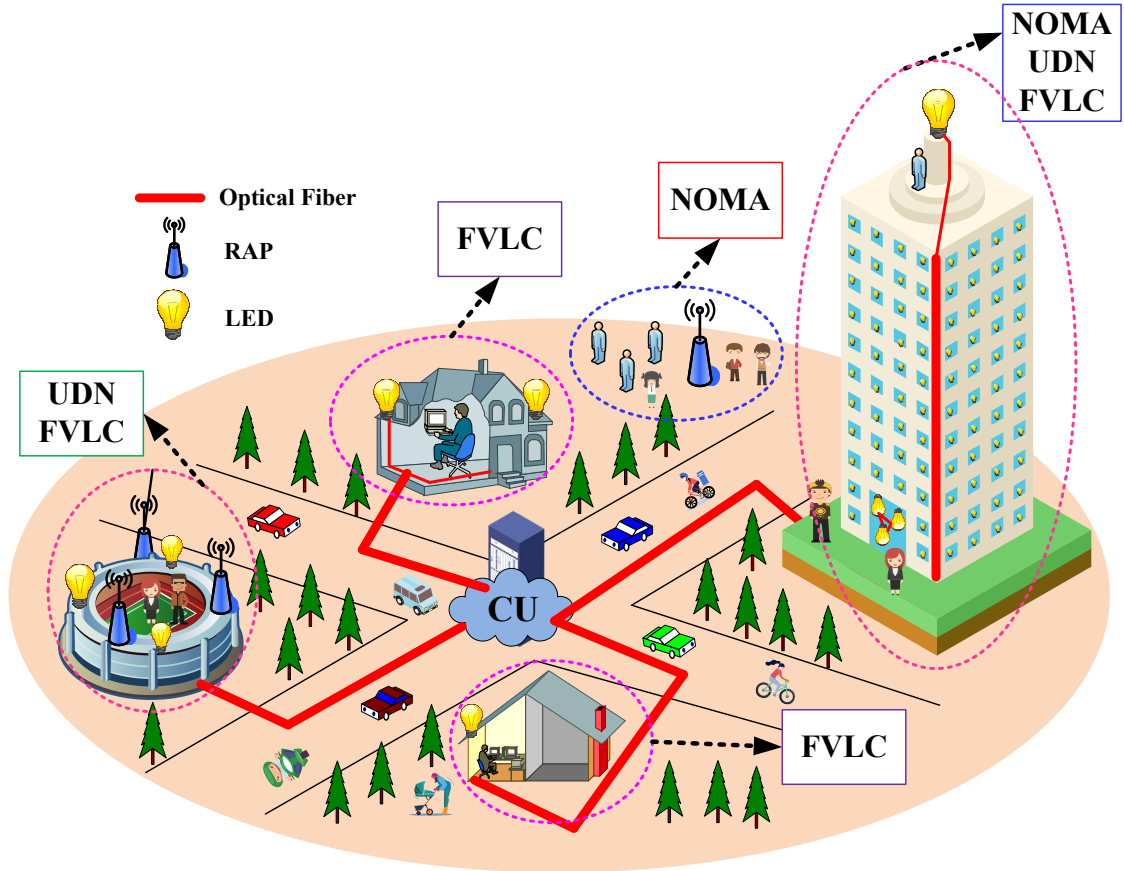}
\caption{System architecture of massively distributed access system with advanced technologies.}\label{fig:scenario}
\end{figure*}

\begin{enumerate}[(i)]
\item 6G should be a ubiquitous and integrated network with broader and deeper coverage, from land, ocean to space. In other words, integrated ground-water-air-space communication will be a major research direction in 6G. Some key technologies which are expected to support 6G networks include visible light communication (VLC), NOMA, unmanned aerial vehicle (UAV) \cite{Jacob2020intelligent,Jacob2020bidirectional}, underwater communications and so on.
\item 6G is expected to work on a higher frequency to achieve wider bandwidth, such as mmWave, Terahertz, visible lightin order to attain 10 to 100 times data rates increase. However, a main challenge is the significant signal attenuation and pass loss in high frequency environments. 
\item 6G will be a personalized intelligent network. By widely using artificial intelligence (AI) technologies \cite{chen2019artificial}, 6G networks will realize virtualized personal mobile communications and evolve into a centric system from three different dimensions, i.e., user centric, data centric and content centric. This is in sharp contrast to the traditional one-dimensional network centric approach. 
\item 6G network will have an endogenous security scheme. By introducing trust and safety mechanisms, 6G will enjoy the capability of self-awareness, real-time dynamic analysis, and adaptive risk management, helping realize cyber space security. On the other hand, block chain and quantum key distribution in 6G will further enhance the system security and privacy.
\item 6G is anticipated to generate massive data through Internet of everything (IoE). In conjunction with other new technologies, such as cloud computing, edge computing, block chain, AI, etc., 6G will realize intelligent everything. In short, 6G will finally support a ubiquitous intelligent mobile world.
\end{enumerate}

\section{Massively Distributed Access System Architecture}

As shown in Fig. \ref{fig:scenario}, typical operation scenarios of massively distributed access systems include dense urban areas such as downtown areas of large cities, sport venues such as football stadiums, or industrial environments such as big manufacturing plants. These scenarios need to support massive number of concurrent connected devices with ultra-high density, which can benefit from the deployment of geographically distributed antennas or access points (APs) \cite{yu2020massively}.

Such a network can be constructed and deployed by taking advantage of existing infrastructure. The optical front-hauls can be implemented by utilizing existing optical fiber infrastructure.
For example, in a dense urban area, huge number of distributed antennas or APs can be installed on the windows, external walls, and roofs of high-rise buildings by using the existing optical network in the building as optical front-hauls. Similarly, antennas or APs can be installed, for example, circling around the stands of a stadium, or along the shelves or cable conduits in an industrial plant.

Deploying a massive number of spatially distributed APs brings the transmitted signal sources much closer to the users. Such an approach will be particularly beneficial for millimeter wave (mm-wave) communications, where the signal transmission range is small but highly directional. Bringing APs closer to the users can improve quality-of-service (QoS) in terms of coverage, interference management and latency.

\subsection{Fiber-based Visible Light Communication}

Either wireless or optical front-hauls can be used to connect the spatially distributed APs or antennas to a central unit (CU) \cite{yu2019energy}. Wireless front-hauls are easy to setup, but are vulnerable to harsh wireless propagation environments, especially in dense urban areas where massively distributed antennas are needed. If the cell size is much smaller than the AP-CU coverage, then the communication bottleneck may shift from the wireless access links between AP and users to the wireless front-hauls. In addition, the mutual interference between wireless access links and wireless front-hauls may adversely affect the performance of integrated networks with in-band wireless front-hauls.

Optical front-hauls, on the other hand, have extremely large bandwidth (thus high data rates), reliability, and security. Employing optical front-hauls requires an optical infrastructure, the constructions of which may be costly and time consuming. However, once the infrastructure is place, it can serve many different applications for decades to come. In addition, optical front-hauls can be constructed by taking advantage of existing optical infrastructure pre-wired in buildings and industrial plants. Thus optical front-haul is a good long-term investment.

The combination of optical fiber front-hauls and VLC access links forms the so-called fiber-based visible light communication (FVLC), which enables us to push the VLC access links as close to the users as possible. In FVLC systems, each light emitting diode (LED) can be considered as an AP for VLC access. Due to the short distances between LEDs and users, the VLC access links can have better QoS performance for more users. Due to the combination of fiber front-hauls and VLC access links, it is thus critical to quantitatively identify the impacts of optical fiber front-hauls over the entire network by considering the complex interactions between the wireless and optical domains.

\subsection{Ultra-dense Network}

Ultra-dense network 
is one of the most promising technologies to bridge the gap between user demands and spectrum resources \cite{kamel2016ultra}. A UDN is defined as a network with the spatial density of cells much larger than that of active users. A large number of low power radio access points (RAPs) are densely deployed in a UDN, each providing coverage over a very small area with radius in the order of tens of meters. The RAP density in a UDN can be as high as 1000 RAPs/km$^2$. The densely deployed RAPs bring cells closer to the users and hence the quality of service (QoS) of each user can be significantly improved. In some works, the spectrum and energy efficiencies of UDN are studied under different deployment strategies, such as the densification of classic macro-cell base stations (BSs), ultra-dense indoor femtocell BSs, and outdoor distributed antenna systems.

In addition, a new kind of UDN scheme called user-centric UDN (UUDN) has been proposed, where multiple RAPs can be grouped together to serve a single user \cite{yu2020massively}. UUDN breaks the traditional cellular-centric architecture. The services and functions of RAPs are implemented solely based on the requirements of each user. Compared to cellular-centric architecture, the user-centric architecture can improve spectrum utilization, interference management, mobility management and effectively eliminate the cell boundaries (thus it is also called cell-free massive MIMO).
One of the main challenges faced by the design of UDN and UUDN is the high energy consumption due to high RAP density. In addition, densely deployed RAPs may increase the interference level across the entire network.

\subsection{Grant-free Non-orthogonal Multiple Access}
With the explosive growth and deployment of wireless communication devices in factory automation, smart city,  autonomous driving, smart health care, etc.
massive machine-type communication
(mMTC) is considered as a representative service category in B5G and 6G mobile communication networks\cite{7973146}. Massive connectivity and low latency
are the two main requirements of mMTC. Unlike conventional human-centric communications, only a small fraction of large-scale potential users are active for exchange of short-packet data.

In LTE and 5G NR, single carrier frequency division multiple access (SC-FDMA) and orthogonal frequency division multiple access (OFDMA) have been adopted in uplink and downlink, respectively. Orthogonal time/frequency resources are allocated to different users. Such a scheme may not be able to support highly massive connections due to limited time/frequency resources. In addition, the significant signaling overhead and
excessive latency caused by complicated handshaking and scheduling procedure
are highly inefficient for the exchange of sporadic short packets. Recently, grant-free NOMA schemes have been considered as a compelling alternative. In grant-free NOMA, multiple users
transmit short-packet data through the same time/frequency resource
without the granting procedure. Consequently, grant-free NOMA can achieve excellent performance in terms of  resource utilization and
latency/signaling overhead.

\begin{figure*}[t]
\centering
\includegraphics[width=0.9\textwidth]{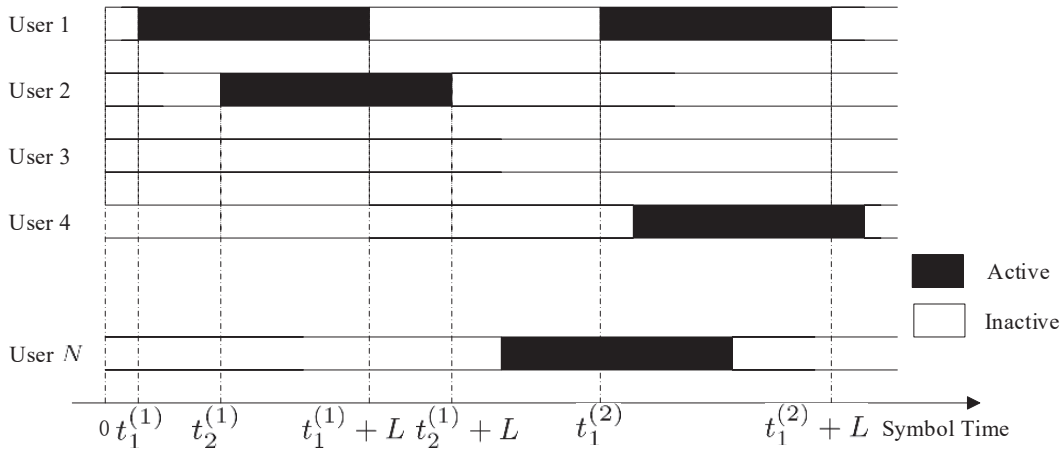}
\caption{Time-domain structure of grant-free NOMA scheme.}\label{IoT}
\end{figure*}

Major challenges of signal detection in mMTC include active use detection (AUD), channel estimation (CE and multi-user data detection. As illustrated in Fig. \ref{IoT}, different active users transmit $L$-length packets
at any time. In addition, the multi-user interference potentially limit the system performance.

An efficient way for AUD and CE is to formulate the two problems jointly as a sparse signal recovery problem and then attack it by the compressed sensing theory. In \cite{8323218}, approximate message passing (AMP) algorithm is used to detect active users and estimate their channel, where Gaussian matrix is considered as a pilot matrix. From another perspective, the AUD and CE may be formulated as a linear inverse problem. However, there are
some disadvantages of the existing detection algorithms with random matrices: 1) there is no
efficient algorithm to check the restricted isometry
property (RIP) of a pilot matrix; 2) the current detection algorithms mostly require significant amount of pilots and
large storage space. Therefore, how to design a deterministic pilot matrix and the corresponding detection method is a research avenue to be explored. For example, the Reed-Muller can be utilized for the design of AUD and CE.

\begin{figure*}[t]
\centering
\includegraphics[width=0.9\textwidth]{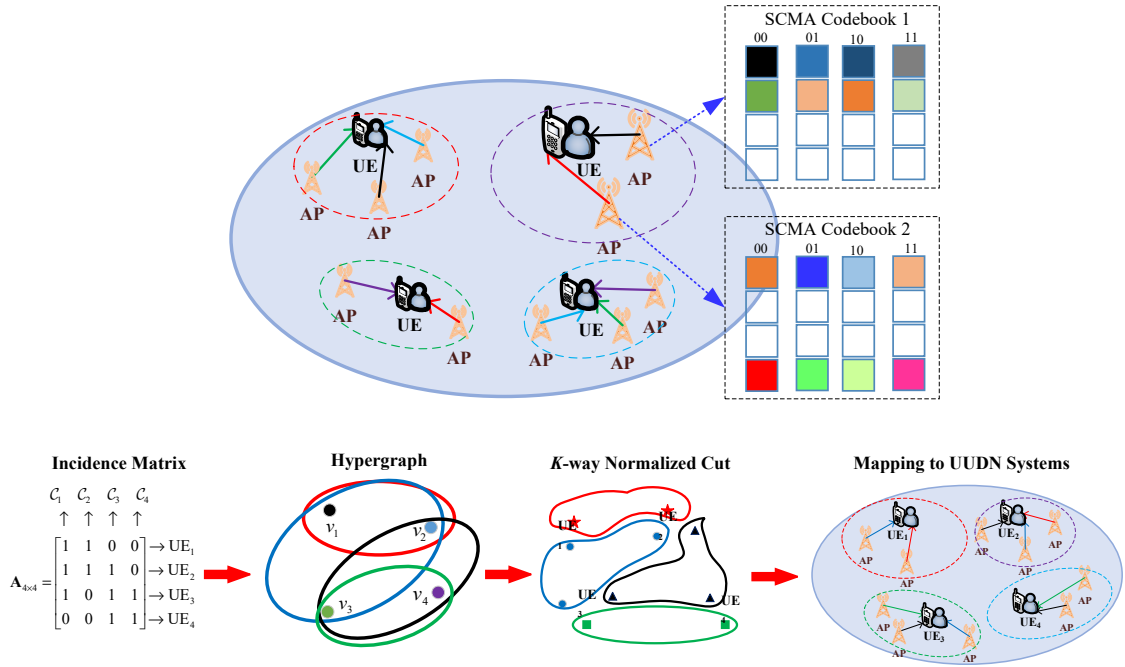}
\caption{An example of SCMA codebook allocation based application in user-centric ultra-dense networks.}\label{fig:UUDNSCMA}
\end{figure*}

For multi-user detection (MUD) in mMTC, a low-complexity detection algorithm based on alternating minimization is proposed in \cite{8642914}. Their proposed algorithm is specifically designed to avoid any
matrix inversion and any computations of the Gram matrix at
the receiver for large-scale systems. In addition, an estimator based bilinear generalized approximate message passing (BiG-AMP) and loop belief propagation (LBP) have been proposed for joint active user, channel and data estimation in \cite{9138688}. The obtained results show that the proposed algorithm has better estimation performance with less pilots.

%\vspace{-4.1mm}
\section{Advances of Sparse Code Multiple Access}

In order to further improve the performance of NOMA,
a code-domain NOMA, namely sparse code multiple access
(SCMA), has emerged recently, in which multiple users can be served simultaneously by employing different codebooks. At the receiver side, a multi-user detector based on message passing algorithm (MPA) is adopted to carry out multiuser detection. Taking advantage of the sparsity of codebooks, MPA detector enjoys significantly lower complexity compared to the maximum likelihood (ML) detector. A practical way to transmit SCMA is to divide all the users into several groups with a small number in each group. In each group, every layer or user has its dedicated codebook. Thus, codebook design plays an important role for optimized SCMA systems. What is more, SCMA is considered as a typical and important grant-free non-orthogonal multiple access. In the sequel, we provide an overview of SCMA:

\begin{enumerate}[(A)]
%\subsection{Codebook Design}
\item \textbf{Codebook Design}:
Extensive research works on SCMA codebook design have
been presented in the literatures. Compared to its original version, called low density signature (LDS), SCMA outperforms due to the shaping gain of multi-dimensional codebooks. Inspired by this, SCMA codebooks based on star quadrature amplitude
modulation (Star-QAM) constellations and constellation rotation
have been proposed \cite{yu2018design}. So far, the optimal SCMA codebook design remains largely open.
%
%
%
%\subsection{Performance Analysis}
\item \textbf{Performance Analysis}:
The performance analysis of SCMA has
drawn a significant amount of research attention. The upper bound
of symbol error rate (SER) of SCMA system was derived based
on the pairwise error probability (PEP) in 2017. Furthermore, a lower
bound of SER has been obtained when the overloading
of the system is small. Subsequently, a theoretical expression for the
BER performance over additive white Gaussian noise (AWGN)
channel has been derived, based on the distribution of the phase
angle in constellations. As an extension, the authors in \cite{yu2018design} studied the BER performance over Rayleigh fading channels.
However, most of the existing analytical
works are not able to provide closed-form expressions of BER
associated to the adopted SCMA codebooks. Thus, it is an open research to study BER performance analysis limited to the actual SCMA codebooks.
%
%
%\subsection{Combinations and Applications}
%
\item \textbf{Combinations and Applications}: As shown in Fig. \ref{fig:UUDNSCMA}, in a massive access scenario, how to integrate UDN and SCMA is a critical research problem. It is interesting to study the UUDN system performance optimization with SCMA using hypergraph theory. Those APs serving UEs in the same cluster
share the same codebooks. On the other hand, when their distances are very large,
those APs in the same AP groups (APGs) may share different codebooks. This can help reduce the interference as much as possible, and
maximize the sum rate of the UUDN network.
\end{enumerate}

\section{Future Directions of SCMA}
\begin{figure*}[t]
\centering
\includegraphics[width=0.7\textwidth]{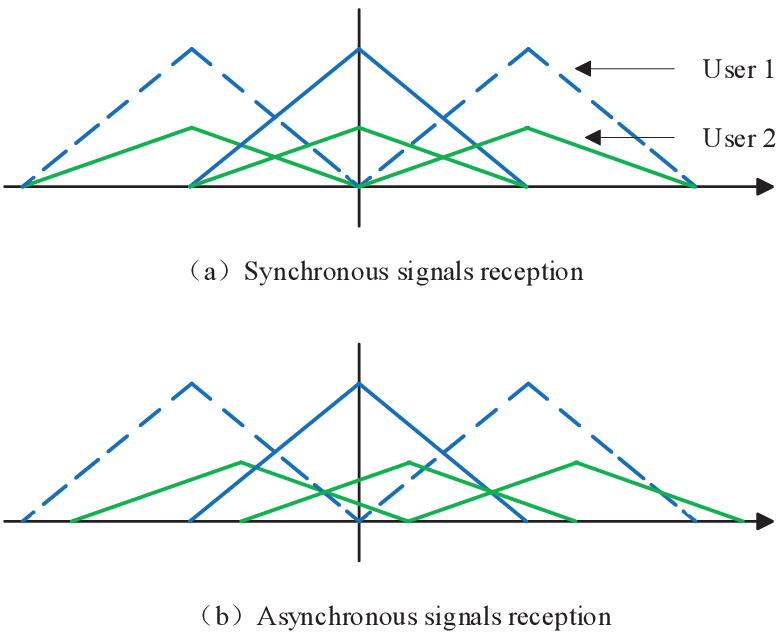}
\caption{A signal reception example of uplink SCMA systems.}\label{Asy}
\end{figure*}

SCMA is a code domain NOMA, which can greatly improve the wireless communication networks' capacity with different overloading factors. However, there are still some challenges, especially for 6G networks. Some future directions are summarized as follows.
In this section, we present a few promising directions on future SCMA research. One can see that many interesting SCMA problems are open. Attacking these problems may be of interest for massive access in 6G. 
\begin{enumerate}
\item \textbf{Optimal SCMA Codebook Design}: The existing SCMA codebook design largely relies on a common multidimensional mother constellation with which multiple codebooks are generated. So far, the design of optimal SCMA codebooks is an open problem and it is unknown how close the current SCMA codebooks to the optimal ones. It is thus imperative to investigate the fundamental limits of SCMA codebooks from information theoretic viewpoints, which may serve as the design guidelines towards the optimal design.
\item \textbf{Large-scale SCMA System Design}: A SCMA scheme with massive connectivity and higher data rate should be developed. However, the lack of scalability in SCMA codebook design limits the application of SCMA in large-scale systems. Albeit numerous research attempts have been carried out, they mostly focus on small-scale system settings (e.g.,  $4 \times 6$ and $5\times 10$). As such, the room for performance improvement is relatively small. Hence, it is interesting to look for new design paradigm (e.g., scalable codebook design, new transmission strategies, and/or low-complexity message passing algorithms at the receiver) for significant performance boost. 
\item \textbf{New Coded SCMA}: Whilst the sparsity of SCMA can be exploited for efficient message passing decoding, it also results in limited diversity order which is a fundamental bottleneck for significant error rate performance enhancement.  Motivated by this issue, apart from adopting a strong channel code, it is worthy investigating spatial coupling aided SCMA (SC-SCMA), where SC is an effective approach for improved BP decoding threshold in coding theory. Further research is needed in understanding the performances of SC-SCMA under different channel conditions. Moreover, when low-density parity-check (LDPC) code is employed, one may also explore the joint sparsity of both LDPC and SCMA with which new message passing algorithms may be devised.
\item \textbf{Cross-layer Optimisation of SCMA/OMA System}: In future mobile communications, SCMA may co-exist with other orthogonal multiple access (OMA) schemes (e.g., OFDMA) to serve a vast range of data services with different quality-of-service (QoS) requirements. As SCMA can support more communication links simultaneously, ultra-low latency is expected. On the other hand, while OMA systems can enjoy better error rate performances, they may suffer from larger collision rates in a random access system.  Hence, cross-layer optimisation between PHY and medium access control (MAC) may be necessary with respect to specific QoS requirements on latency, reliability, energy- and spectrum- efficiencies.
\item \textbf{SCMA with Synchronization Errors}:
A signal reception example of uplink SCMA systems shown in Fig. \ref{Asy}, there are exist different time offset when signals from different devices arriving at the receiver asynchronously. This is because that devices are geographically distributed and signals from different devices begin to propagate at any time. Most of the existing works on SCMA are based on the assumption of perfect synchronization or timing advance mechanism. For small number of UEs, OFDM signal with cyclic prefix (CP) can be used to eliminate synchronization errors. However, when the number of UEs is large, the CP length is very long, which is not practical in the massively distributed access scenario. Thus, As a result, designing a detection algorithm for SCMA with synchronization errors, including the time-offset and the frequency-offset is necessary. Moreover, these estimation errors are often unknown, which is more challenging.
\item \textbf{SCMA for High Mobility Communications}: The existing SCMA studies are mostly focused on its application for massive machine-type communications with low mobility. With the rapid evolution of intelligent transportation systems \cite{khan2020an}, the prevalent vision is that connected autonomous vehicles (CAVs) will be seen everywhere. Challenges may arise when a CAV moving in high speeds (e.g., larger than 500 km/h) carries out information exchange using SCMA. In this case, innovative solutions are needed to maintain an excellent reception quality. A promising direction is to study the integration of SCMA with orthogonal-time-frequency-space (OTFS), which is a modulation technique with displays superior performance in high Doppler environments. It is expected that OTFS-SCMA system will provide massive connectivity in the high speed vehicle-to-everything (V2X) scenario.
\end{enumerate}

\section{Conclusions}

In this article, we have proposed to use SCMA to support massively distributed access systems in 6G for faster, more reliable, and more efficient massive access.
To this end, we first provided a brief overview on mobile evolution from 1G to 5G. Then, we presented our visions on 6G wireless communication networks.
We have introduced the MDAS architecture in details with the developments of FVLC, UDN and NOMA. Among these three techniques, SCMA is shown to be an excellent candidate for reshaping a massively connected multiple access system. Finally, we presented our thoughts on future directions of SCMA.

\end{document}